\documentclass[prl,preprintnumbers,superscriptaddress,amsmath,amssymb,floatfix]{revtex4}

\usepackage{graphicx}
\usepackage{epsfig}
\usepackage{esint}

\begin{document}

\title{Vacancies, disorder-induced smearing of the electronic structure, and its implications for the superconductivity
of anti-perovskite MgC$_{0.93}$Ni$_{2.85}$}

\author{David~Ernsting}
\affiliation{H.H.~Wills Physics Laboratory, University of Bristol, Tyndall Avenue, Bristol, BS8 1TL, United Kingdom}
\author{David~Billington}
\affiliation{Japan Synchrotron Radiation Research Institute, SPring-8, Sayo 679-5198, Japan}
\author{Thomas~E.~Millichamp}
\affiliation{H.H.~Wills Physics Laboratory, University of Bristol, Tyndall Avenue, Bristol, BS8 1TL, United Kingdom}
\author{Rebecca~A.~Edwards}
\affiliation{H.H.~Wills Physics Laboratory, University of Bristol, Tyndall Avenue, Bristol, BS8 1TL, United Kingdom}
\author{Hazel~A.~Sparkes}
\affiliation{School of Chemistry, University of Bristol, Cantock's Close, Bristol, BS8 1TS, United Kingdom}
\author{Nikolai~D.~Zhigadlo}
\affiliation{Department of Chemistry and Biochemistry, Freiestrasse 3, University of Bern, Bern, Switzerland}
\author{Sean~R.~Giblin}
\affiliation{School of Physics and Astronomy, Cardiff University, Queen's Building, The Parade, Cardiff, CF24 3AA, United Kingdom}
\author{Jonathan~W.~Taylor}
\affiliation{DMSC - European Spallation Source,  Universitetsparken 1, Copenhagen 2100, Denmark}
\author{Jonathan~A.~Duffy}
\affiliation{Department of Physics, University of Warwick, Coventry, CV4 7AL, United Kingdom}
\author{Stephen~B.~Dugdale}
\email{s.b.dugdale@bristol.ac.uk}
\affiliation{H.H.~Wills Physics Laboratory, University of Bristol, Tyndall Avenue, Bristol, BS8 1TL, United Kingdom}

\begin{abstract}
\noindent
The anti-perovskite superconductor MgC$_{0.93}$Ni$_{2.85}$ was studied using high-resolution x-ray Compton scattering
combined with electronic structure calculations.
Compton scattering measurements were used to determine experimentally a Fermi surface that showed good agreement
with that of our supercell calculations, establishing the presence of the predicted hole
and electron Fermi surface sheets.
Our calculations indicate that the Fermi surface is smeared by the disorder due to the presence of vacancies on the C 
and Ni sites, but does not drastically change shape.  The 20\% reduction in the Fermi level density-of-states would lead to 
a significant ($\sim 70\%$) suppression of the superconducting $T_c$ for pair-forming electron-phonon coupling. However, we ascribe 
the observed much smaller $T_c$ reduction at our composition (compared to the stoichiometric compound) to the suppression of 
pair-breaking spin fluctuations. 
\end{abstract}

\maketitle

\section{Introduction}

\noindent
MgCNi$_{3}$ has been the focus of a great deal of interest since the discovery of superconductivity below a
critical temperature of $T_{c}\approx8$~K \cite{he:01}.
The primary cause of this interest was that superconductivity existed at all in the presence of such a
high proportion of Ni, suggesting that the compound may be on the verge of magnetic order and harbour strong
ferromagnetic spin fluctuations which are antagonistic to spin-singlet superconductivity.
This resulted in speculation that MgCNi$_{3}$ might be an unconventional electron-paramagnon superconductor, its 
pairing driven by interactions involving ferromagnetic spin fluctuations,
as speculated for the putative spin-triplet $p$-wave superconductor, Sr$_{2}$RuO$_{4}$ \cite{ishida:98}.
Further interest in this intermetallic compound has stemmed from the fact that it crystallises in the cubic
perovskite structure that is the building block of both Sr$_{2}$RuO$_{4}$ and the layered cuprate
high-temperature superconductors \cite{norman:03}.
The structural similarity that it shares with these compounds led to speculation
that it may provide the link between the unconventional superconductivity exhibited
by high-temperature superconductors, and conventional superconductivity in intermetallic compounds \cite{klimczuk:04}.

Since the initial surge of interest, there followed significant controversy regarding the nature of the pairing responsible for the superconductivity,
with various studies suggesting a possible spin-fluctuation-driven mechanism
\cite{rosner:01,prozorv:03,young:04}, multiband superconductivity \cite{voelker:02}, conventional electron-phonon coupling
\cite{he:01,lin:03,shan:03a,shan:03b,ignatov:03,walte:04} including a C isotope effect \cite{klimczuk:04}, and some even providing evidence
in support of both \cite{singer:01,mao:03}.
The first calculations of the electron-phonon coupling constant, $\lambda$, indicated that it was relatively large (Dugdale and Jarlborg
calculated $\lambda\approx0.9$ \cite{dugdale:01}, and Singh and Mazin suggested that low-frequency rotational modes of the Ni octahedra
could contribute a $\lambda$ of about 3 on the lighter FS sheet \cite{singh:01}). Subsequent calculations indicated that the overall
$\lambda$ could be as large as 1.51 due to anharmonic effects \cite{ignatov:03}.
The Sommerfeld parameter is also consistent with a $\lambda$ of 1.45, and the jump in the specific heat at the
superconducting transition was found to be strongly enhanced compared to the BCS value,
and interpreted as being due to strong electron-phonon coupling by W\"alte {\it et al.} \cite{walte:04}.
Furthermore, this strong-coupling version of superconductivity is supported by more recent experiments \cite{diener:09,pribulova:11} and theory
\cite{szczesniak:15}.
Contrary to this, the most recent study, based upon London penetration depth measurements that probe the underlying superconducting gap structure,
concluded that MgCNi$_{3}$ is a conventional, $s$-wave, weak-coupling superconductor \cite{gordon:13}.
However, their observations are also consistent with the antagonistic coexistence of relatively strong pair-forming electron-phonon coupling and pair-breaking
electron-paramagnon coupling due to the strong spin fluctuations expected in a material so close to ferromagnetism \cite{shan:05}, the combination of which
would conspire to suppress the critical temperature \cite{daams:81}.
Further clarification of the nature of the superconductivity requires detailed experimental
studies of the electronic structure.

The only direct electronic structure studies have been via x-ray emission, electron photoemission,
and x-ray absorption spectroscopy experiments \cite{shein:02,kim:02}, which probe the total and partial
density-of-states (DOS). The samples studied in those experiments were polycrystalline, some of which were slightly
deficient in C or doped with Co, with the highest $T_{c}$ observed close to perfect stoichiometry \cite{amos:02}.
Nevertheless, these experiments were consistent with the calculated electronic
structure and, significantly, verified the existence of a peak in the DOS just below the
Fermi energy, $E_{\rm F}$. The height of this peak was, however, found to be greatly suppressed with respect
to the existing predictions \cite{dugdale:01,singh:01,rosner:01,shim:01,shein:02}.
Clearly an experimental measurement of the Fermi surface of MgCNi$_{3}$ is highly desirable \cite{dugdale:16}, 
and the recent availability of suitable single crystals has made this possible.
Here, we have utilised high-resolution x-ray Compton scattering to determine
the bulk Fermi surface and electronic structure of a C- and Ni-deficient single crystal with composition MgC$_{0.93}$Ni$_{2.85}$. 
Compton scattering is not restricted by short electronic mean-free-paths, does not rely on a cleanly cleaved surface, and probes
the bulk of the material \cite{dugdaleltp}.
These measurements are compared with state-of-the-art electronic structure calculations to help interpret the
data and understand the effect of disorder-induced smearing of the quasiparticle states on the bulk electronic
structure.

\section{Results and Discussion}

The stoichiometry of the samples was determined by x-ray diffraction (XRD) on one of the smaller
single crystals in the growth batch, and found to be Mg$_{1.00\pm0.01}$C$_{0.93\pm0.01}$Ni$_{2.85\pm0.03}$. Samples in this
batch typically had $T_{c}\approx 6.5$~K.

In order to understand the experimental results and anticipate the effects of disorder from the presence of vacancies,
we first reproduce what is known from earlier calculations about the
electronic structure of the stoichiometric compound, before moving on to that of the non-stoichiometric
compound.
This will provide a basis for understanding the electronic structure of MgC$_{0.93}$Ni$_{2.85}$ relative to MgCNi$_{3}$.

\subsection{Stoichiometric electronic structure calculations}

There is consensus among previous electronic structure calculations about the electronic structure of stoichiometric
MgCNi$_{3}$; there is a large peak in the DOS situated just below $E_{\rm F}$, comprised of mainly
Ni $3d$ and C $2p$ states, that results from a van Hove singularity (vHs) caused by a very flat,
high-mass band around the M-points of the simple cubic Brillouin zone 
\cite{szajek:01,dugdale:01,singh:01,rosner:01,shim:01,shein:02}.
Both this high-mass band (band 1) and a second, much lighter band (band 2)
cross $E_{\rm F}$, giving rise to hole and electron Fermi surface sheets, respectively.
As the unit cell has an even number of electrons, MgCNi$_{3}$ is a compensated metal
and these sheets contain equal numbers of holes and electrons.
The calculated electronic structure of the stoichiometric compound 
was able to reproduce the experimentally observed temperature dependence of
the Hall coefficient \cite{li:01}, normal state resistance \cite{kumary:02}, and thermopower \cite{li:02}.
From this electronic structure, W\"alte {\it et al.} decomposed the electron-phonon coupling between the hole and electron FS sheets,
giving $\lambda_{1}\approx1.74$--1.78 and $\lambda_{2}\approx2.20$--2.42, respectively \cite{walte:04}. 

The electronic structure of stoichiometric MgCNi$_{3}$ calculated in the present study is essentially the same
as the previous studies: Figs.~\ref{calc}a, b and c show the DOS and Fermi surface sheets, respectively.
The DOS at the Fermi energy, $N(E_{\rm F})=5.63$~states~(eV~f.u.)$^{-1}$, is in good agreement with the previous
calculations, and is dominated by band 1, which contributes $83\%$ of $N(E_{\rm F})$.
As in the previous studies, two bands cross $E_{\rm F}$ creating a hole sheet (band 1) and an electron sheet (band 2).
The band 1 hole sheet consists of an X-centred shell-like feature and eight ovoids located between the $\Gamma$- and R-points,
whilst the electron sheet from band 2 consists of a $\Gamma$-centred distorted octahedron and a jungle-gym connecting the R- and M-points
(around the edges of the simple cubic Brillouin zone).

\subsection{Non-stoichiometric electronic structure calculations}

Calculations for the non-stoichiometric compound were performed with various methods (see {\it Methods})
in order to attempt to treat the effect of disorder on the electronic structure.

\begin{figure}[t!]
%\centerline{\includegraphics[width=0.8\linewidth]{Figure-1_Dugdale.eps}}
\centerline{\includegraphics[width=0.8\linewidth]{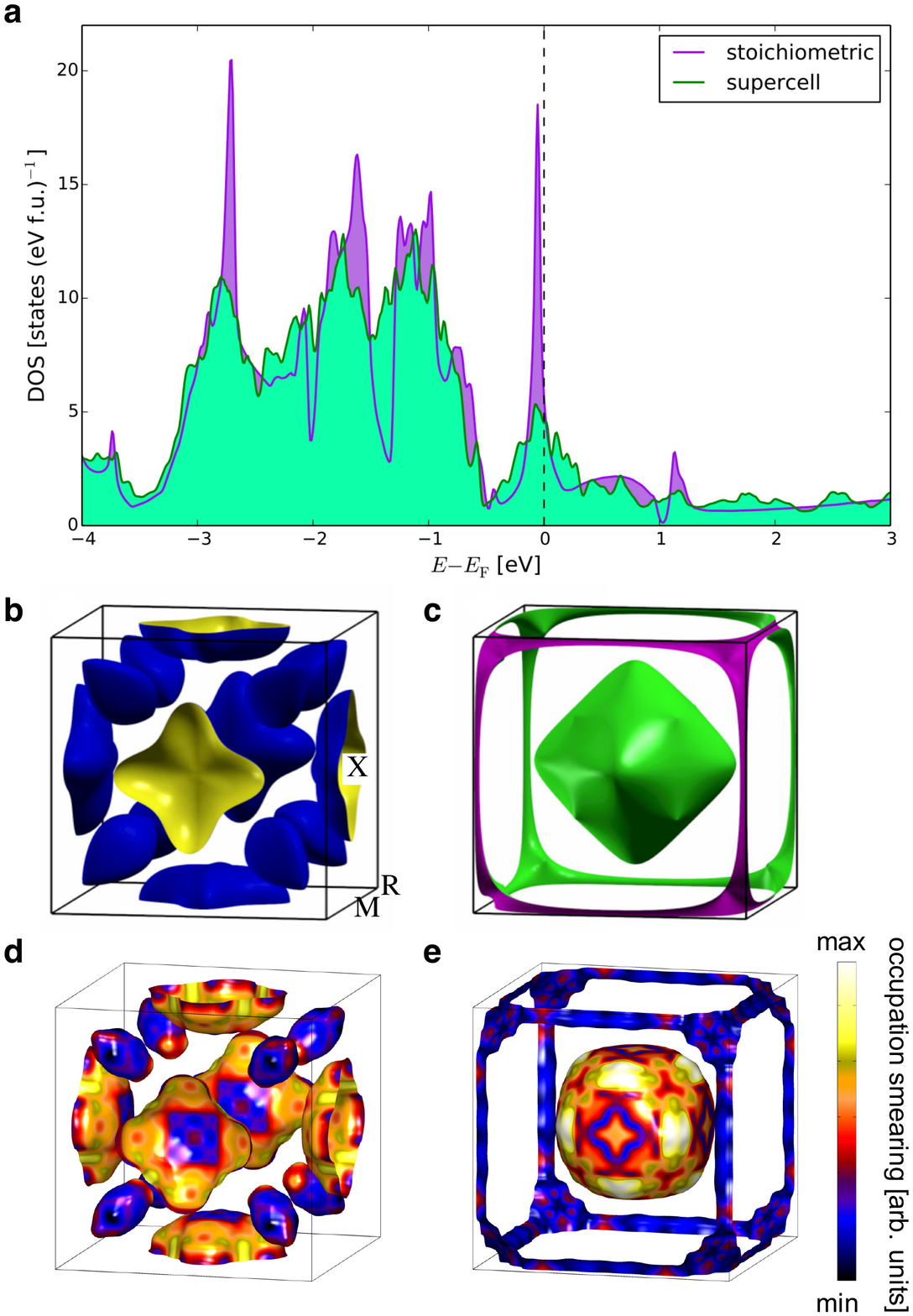}}
\caption{{\bf Calculated electronic structure of stoichiometric and non-stoichiometric MgCNi$_{3}$.}
{\bf a}, Comparison of the total DOS for the stoichiometric (purple) and supercell configurational
average (green) calculations.
Calculated stoichiometric Fermi surface from bands 1 and 2 are shown in {\bf b} and {\bf c}, respectively.
Calculated MgC$_{0.875}$Ni$_{2.875}$ supercell configurationally averaged Fermi surface
of bands 1 and 2 are shown in {\bf d} and {\bf e}, respectively. The colours on the surfaces represent the occupation smearing of
the configurational average, and are given by the standard deviation of the occupation density
at the Fermi surface.
The high-symmetry points of the simple cubic Brillouin zone are labelled in {\bf b}.}
\label{calc}
\end{figure}

Fig.~\ref{calc}a also shows the calculated DOS for a
$2\times2\times2$ supercell configurational average (with one C and one Ni atom removed to give an effective
stoichiometry of MgC$_{0.875}$Ni$_{2.875}$). 
Evaluating the momentum density from such a supercell is already computationally demanding,
and obtaining a configurational average from a series of larger supercells was not practical. However, the 
DOS from a $3\times3\times3$ supercell was found to be very similar to, and slightly smoother than the smaller 
supercell, as might be anticipated. 
The supercell configurational average predicts a very similar DOS (particularly near $E_{\rm F}$),
but the large peak just below $E_{\rm F}$ exhibits a broadening and a significant height reduction,
presumably due to a smearing of the vHs in comparison to the stoichiometric calculation.
Interestingly, these calculations show no clear shift of $E_{\rm F}$ relative to the band manifold (as is
predicted by a so-called `virtual crystal approximation' (VCA) \cite{nordheim:31} calculation), in agreement
with the electronic structure observed by the spectroscopy experiments on polycrystals with C vacancies \cite{shein:02,kim:02},
and previously predicted to appear in `coherent potential approximation' (CPA) calculations with C vacancies \cite{shan:03a}.
Because of the proximity of the Fermi level to the diminished vHs, $N(E_{\rm F})$ is reduced to $4.52$~states~(eV~f.u.)$^{-1}$.
This reduction in $N(E_{\rm F})$ compared with the stoichiometric compound would indicate a reduced propensity for
magnetism and spin fluctuations in the presence of C and Ni vacancies (in accordance with the lack of evidence reported
for spin fluctuations in experiments performed upon single crystals \cite{pribulova:11}, compared to near stoichiometric
polycrystals \cite{walte:04}).
The reduction in $N(E_{\rm F})$ may also explain why single crystals with C and Ni vacancies are observed to
have a lower $T_{c}$ than polycrystals with small amounts of C deficiency \cite{he:01,amos:02,lee:07,gordon:13}.

The configurationally averaged supercell Fermi surfaces are shown in Figs.~\ref{calc}d and e.
The supercell calculations predict the same number of Fermi surface sheets as in the stoichiometric
compound, and they retain the same general shape.
Although there is no longer an even number of electrons per unit cell, the hole
and electron sheets have very similar volumes, as may be expected from the unusual temperature
dependence of the Hall coefficient and thermopower (measured in polycrystals)
\cite{li:01,li:02}, explained by hole and electron sheets with almost equal volumes \cite{singh:01,kumary:02}.
The colours shown on the surfaces represent one standard deviation of the occupation densities from the
individual supercell calculations relative to the average, at the configurationally averaged Fermi surface.
This provides an indication of the size of the disorder smearing in ${\bf k}$-space, and
suggests that the $\Gamma$-centred electron and X-centred hole sheets are more sensitive to
the disorder than the outer electron jungle-gym and the hole ovoids between the $\Gamma$- and R-points.
However, the smearing is not large, with well-defined darker regions
showing negligible variation, and even the lighter regions indicate a smearing of less than $3\%$ of the Brillouin zone.
It should be emphasised that this characterisation of the smearing is likely to be sensitive to the supercell size and stoichiometry.

\subsection{Non-stoichiometric electronic structure determined from Compton scattering}

\begin{figure}[t!]
%\centerline{\includegraphics[angle=90,width=0.8\linewidth]{Figure-2_Dugdale.eps}}
\centerline{\includegraphics[angle=90,width=0.8\linewidth]{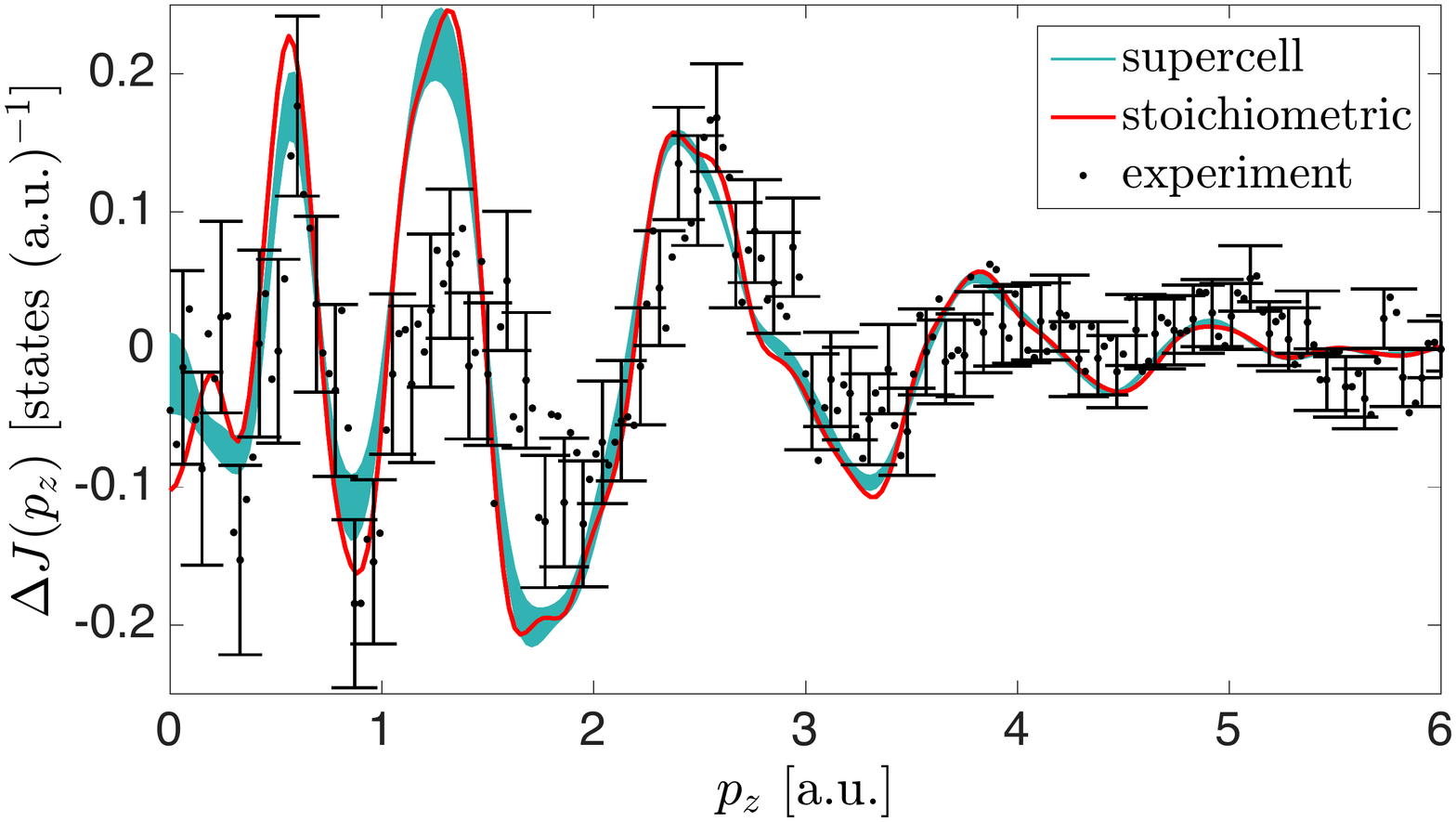}}
\caption{{\bf Comparison between the experimental and calculated Compton profiles.}
Difference, $\Delta J(p_{z})$, between Compton profiles with scattering vectors along the $[110]$ and $[100]$
directions for the experimental profiles (black dots), and those of the stoichiometric (red line) and supercell
configurational average (blue line) calculations.
The thickness of the blue line indicates the variation amongst the different supercell configurations and is two
standard deviations wide. 
All of the calculations have been convoluted with a one-dimensional Gaussian with a full-width-at-half-maximum of 0.12~a.u.
to approximate the experimental resolution. For clarity, the error bars are only plotted for every third data point and indicate statistical errors of one standard deviation.}
\label{dirdiff}
\end{figure}

A Compton profile, $J(p_{z})$, is a one dimensional projection (double integral) of the underlying electron momentum density, 
thus containing information about the 
occupied momentum states and therefore about the Fermi surface \cite{dugdaleltp}.
Six Compton profiles were measured along equally spaced directions between the crystallographic [100] and [110] directions in
the (001) plane (see {\it Methods}).
The greatest directional difference, $\Delta J(p_{z})$, is observed in the difference between the [100]
and [110] directions (Fig.~\ref{dirdiff}),
this being the largest angle between measured directions. Also shown
are the calculated differences between profiles for stoichiometric MgCNi$_{3}$,
and those resulting from the configurational average of the supercell calculations.
These calculated directional differences show reasonable agreement with the experimental differences, with peaks and
troughs in the same places.
Very rarely, the Fermi surface contribution to the directional differences is readily visible in the
data (see, for example, \cite{billington:15}).
However, our calculations confirm that the Fermi surface contribution to the Compton profile anisotropy for
stoichiometric MgCNi$_{3}$ is small, the anisotropy being dominated by the fully occupied bands.
The stoichiometric and supercell calculations are very similar to one another, indicating that the effect of
disorder is subtle in such directional differences. 
However, the size of the anisotropy in both calculations is distinctly overestimated compared to experiment
(particularly the peaks and troughs in the momentum range 1.2 to 3.2~a.u.).
This type of discrepancy is common in electronic structure calculations of the electron momentum density,
and could arise due to limitations in the treatment of electron correlations in the local density approximation
(LDA) \cite{shiotani:93}, although only anisotropic correlation effects would be visible in the directional
differences.
The overestimation of the anisotropy could also arise from self-interaction effects due to the LDA treatment
of the Ni $d$-electrons, as is the case for Cu \cite{kubo:99}.

\begin{figure}[t!]
%\centerline{\includegraphics[width=0.8\linewidth]{Figure-3_Dugdale.eps}}
\centerline{\includegraphics[width=0.8\linewidth]{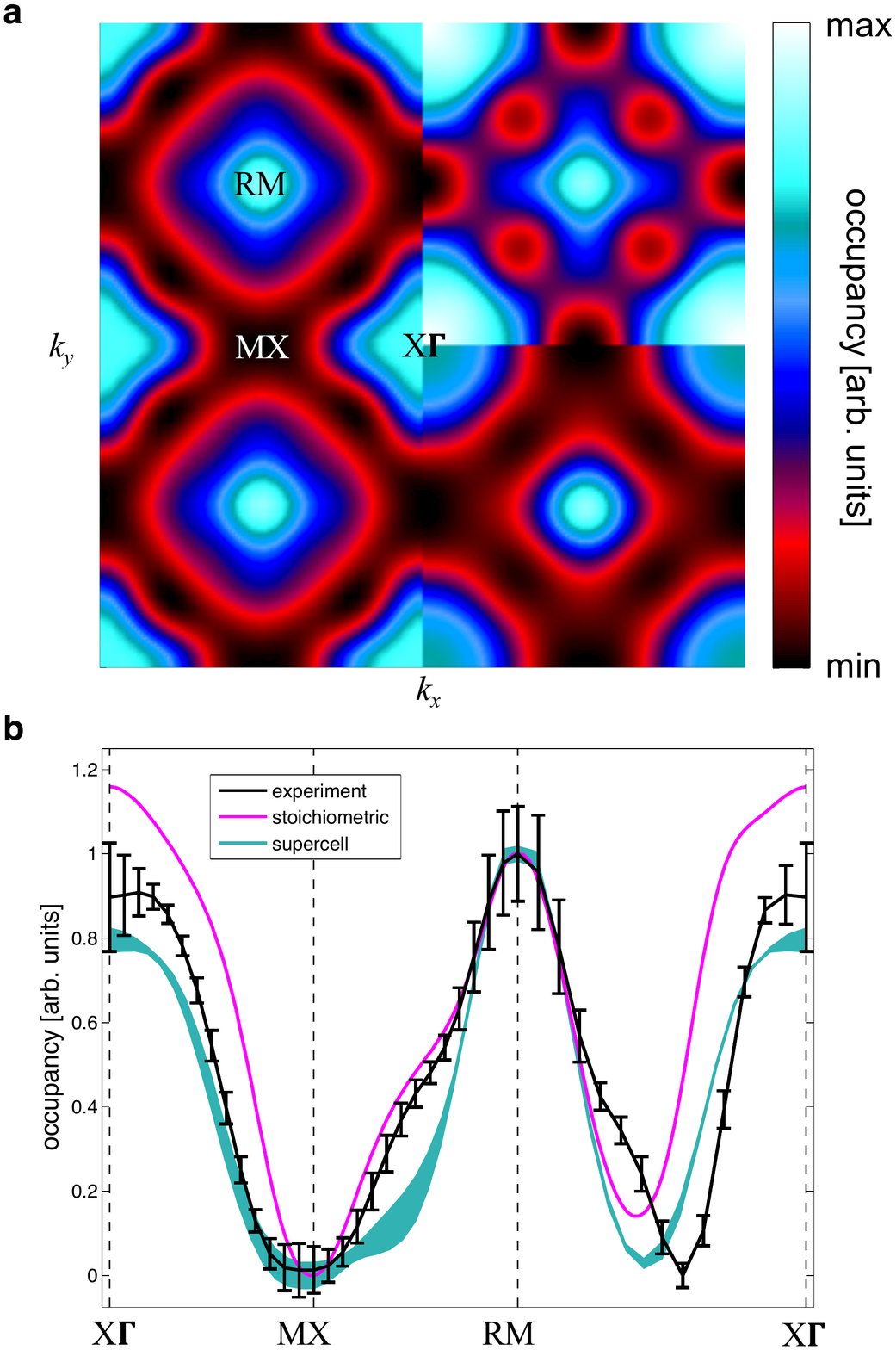}}
\caption{{\bf Comparison between the reconstructed experimental and calculated projected occupation densities.}
{\bf a}, Experimental occupation density (left half), and theoretical ${\bf k}$-space occupation densities,
projected down [001], for the stoichiometric (top right), and supercell configurational average (bottom right) calculations.
The plot is centred at the projected X$\Gamma$-point, and spans two
Brillouin zones. The calculated projected occupation densities have been convoluted with a two-dimensional Gaussian approximating the
experimental resolution function.
The projected high-symmetry points are labelled.
{\bf b}, Cuts through the [001]-projected experimental and theoretical occupation densities along certain
(projected) high-symmetry directions.
The thickness of the line for the supercell configurational average corresponds to two standard deviations and
the experimental error bars indicate statistical errors of one standard deviation.
All distributions have been normalised to unity at the projected RM-point.
}
\label{exp}
\end{figure}

A reconstruction from the 1D Compton profiles to a 2D projection was performed, and the resulting
distribution was  reduced to the first Brillouin zone, following the Lock-Crisp-West prescription \cite{lcw} (see {\it Methods})
such that the occupation density, $n({\bf k})$, projected down the [001] direction was obtained.
Note that, with a sufficiently large number of Compton profiles measured along judiciously chosen directions, it is possible to
reconstruct the three dimensional momentum density and thus the three dimensional Fermi surface \cite{dugdale:06}, however, with
the small single crystal sample used in this study (see {\it Methods}) and a limited period of beamtime, such an approach was not feasible.
Fig.~\ref{exp}a shows the experimental and calculated two-dimensional 
occupation density, projected down [001].
Referring first of all to the stoichiometric calculation (top-right quadrant), the light blue circular regions around the RM points originate
from the projection of the band 2 jungle-gym arms which run the entire length of the Brillouin zone.
The X-centred hole shells (band 1) can be discerned at the projected MX point, and the hole ovoids associated with the same
band can be seen between X$\Gamma$ and RM.
Finally, the structure around X$\Gamma$, which varies from blue to white, originates from the band 2 distorted octahedron,
and its appearance is strongly  influenced by the projection of overlapping projected Fermi surfaces (in particular the hole ovoids).
The experimental occupation density (left) is reproduced rather well by the configurationally averaged supercell (bottom right).
Our KKR-CPA calculations of the projected ${\bf k}$-space occupation density at the supercell composition (not shown, for clarity)
and configurational average supercell predictions of smearing upon the $\Gamma$-centred sheet correctly describe the experimental
occupation density at the projected X$\Gamma$-point.
Moreover, the supercell prediction of a slight thickening of the X-centred hole shells towards the $\Gamma$-point provides a much
better representation of the size of the dip in the experimental occupation density around the projected MX-point.
Between the X$\Gamma$- and RM-points, the position of the dip in the supercell result (caused by the hole ovoids) is slightly closer
to the position of the dip in the experiment than the stoichiometric result, but does not agree completely with experiment, signifying
some slight change in the positions of the crossings of the bands along the $\Gamma$-R direction, or of some additional smearing that
is not reproduced in a supercell of this size.
Figure~\ref{exp}b shows cuts through the $[001]$-projected experimental and calculated occupation densities, that essentially make clear the
observations from the two-dimensional distributions in Fig.~\ref{exp}a.

Our experimental measurement of the Fermi surface of MgC$_{0.93}$Ni$_{2.85}$
reveals both hole and electron sheets of a similar shape to that predicted by LDA calculations for stoichiometric MgCNi$_{3}$.
The presence of C and Ni vacancies makes some parts of the Fermi surface sheet become rather smeared,
but the other regions seem essentially unchanged.
There is some evidence
 that the ovoid hole sheets between the $\Gamma$- and R-points are closer to the $\Gamma$-point than predicted,
but a slight change in their shape (or smearing) could also potentially account for this discrepancy.
That the supercell calculations are not completely able to describe the
occupation density throughout the whole Brillouin zone is perhaps not so surprising given the small size of the supercell.
It might also be important to consider the relaxation of the atomic positions in the presence of the vacancies, as has
been revealed by powder neutron diffraction \cite{amos:02}, but this is beyond the scope of this primarily experimental study.

\section{Conclusions}

We have presented a thorough investigation of the electronic structure and Fermi surface of  MgC$_{0.93}$Ni$_{2.85}$,
through high-resolution x-ray Compton scattering experiments combined with electronic structure
calculations that treat the C and Ni vacancies.
Supercell calculations indicate that the electronic structure, whilst smeared by disorder,
does not drastically change in the presence of vacancies, and confirms the significant reduction in height
of the vHs DOS peak, previously observed from photoemission
experiments in polycrystalline samples with C vacancies \cite{kim:02}.
This reduction in the DOS at the Fermi energy naturally provides an explanation for 
both the reduction in $T_{c}$, and reduced ferromagnetic spin fluctuations
observed in single crystals (which have more C and Ni vacancies than polycrystals).
Our supercell configurational average (at a composition MgC$_{0.875}$Ni$_{2.875}$, which is close to our sample) indicates a
disorder-induced smearing of the stoichiometric Fermi surface.

High-resolution x-ray Compton scattering measurements of the electron momentum density were performed,
from which the projected ${\bf k}$-space occupation density was obtained.
We find good agreement between the experimental occupation density and that predicted by our
configurationally averaged supercell calculation, establishing the
presence of both the hole and electron Fermi surface sheets predicted in calculations and confirming their shapes.
Whilst none of our calculations completely describe the smearing of the occupation density in the
presence of vacancies, we are able to identify significant smearing of the $\Gamma$-centred electron sheet,
and slightly less smearing of the X-centred hole sheets; other parts of the Fermi surface appear to remain sharply defined.

With regards to the superconductivity, the most significant effect of the disorder is the suppression of the vHs
and the reduction in the Fermi level DOS.
If the pairing mechanism is purely electron-phonon, the BCS theory of superconductivity gives $T_{c}\propto\exp(-1/N(E_{\rm F})V)$,
where $V$ is the electron-phonon interaction potential \cite{BCS}.
Given that the experimental superconducting critical temperature extrapolates to a maximum of $T_{c}^{\text{max}}=8.2$~K at perfect
stoichiometry \cite{amos:02}, and assuming that $V$ remains constant, this implies that the reduced $N(E_{\rm F})$ of the
non-stoichiometric sample would give $T_{c}\approx2.4$~K.
Since the experimental superconducting $T_{c}^{\text{exp}}\approx6.5$~K is not suppressed as strongly,
this implies that the pair-breaking effect of spin fluctuations must also be reduced.
Finally, we have unequivocally established that the Fermi surface of MgC$_{0.93}$Ni$_{2.85}$ is qualitatively similar to that predicted
for the stoichiometric compound, and can conclude that the extensive previous theoretical investigations into the properties
of MgCNi$_{3}$ (that do not treat disorder) are based upon an experimentally justified description of its electronic structure.

\section{Methods}

\subsection{Crystal growth}

Single crystals of MgCNi$_{3}$ were grown using a self-flux method within a high-pressure cubic anvil cell
as described in Ref.~\cite{gordon:13}.
A mixture of Mg, C, and Ni powders with a molar ratio of 1:1:3 were placed within
a boron nitride crucible, and placed under a pressure of 3~GPa at room temperature.
The assembly was then heated at constant pressure to above 1600$^{\circ}$C for one hour,
before being slowly cooled to room temperature.
The resultant growth lump was crushed, and single crystal samples were mechanically extracted.

The samples were characterised by magnetic susceptibility experiments
and were found to have a superconducting transition temperature of $T_{c}\approx6.5$~K, with a narrow spread,
presumably due to small variations in Ni and C content throughout the batch.
The observed $T_{c}$ agrees with previous measurements; it is slightly lower than the $T_{c}=6.7$~K observed
for MgCNi$_{2.8}$ single crystals grown at higher pressures \cite{lee:07}, and lower than the $T_{c}=7.3$~K observed
in almost stoichiometric polycrystals \cite{amos:02}.
The stoichiometry of the samples was determined by x-ray diffraction (XRD) experiments upon one of the small
samples in the batch.
The stoichiometry was found to be MgC$_{0.93}$Ni$_{2.85}$ (with an error of $\pm0.01$ for Mg and C,
and $\pm0.03$ for Ni), with a lattice constant of $a=3.8008(2)$~\AA~at $T=100$~K.
These results are consistent with the stoichiometry of MgC$_{0.92}$Ni$_{2.88}$ reported for other single crystals
grown by the same method, but our lattice constant differs from their value of $a=3.7913$~\AA~\cite{gordon:13}.
This indicates a relatively large variation in lattice constant for samples grown by the same method,
with effectively the same stoichiometry, and suggests a potentially significant uncertainty in lattice
constant for different samples throughout the growth batch used in this study.
Interestingly, the transition temperature is higher than the $T_{c}\approx4.5$~K expected for polycrystals with
the C$_{0.93}$ site occupation and a fully occupied Ni site \cite{amos:02}.
The single crystal used in the Compton scattering experiment had the dimensions of approximately $1.0\times0.5\times0.4$~mm$^{3}$.

\subsection{Computational details}

First-principles electronic structure calculations of the stoichiometric and non-stoichiometric compounds were performed.
The highly-accurate full-potential \textsc{Elk} code \cite{elk} with an augmented plane-wave plus
local orbital basis was used for the majority of the calculations.
Within the \textsc{Elk} code, virtual crystal approximation (VCA) \cite{nordheim:31} and supercell calculations were
employed to simulate the effect of vacancies on the electronic structure.
As supercell calculations explicitly include any effects due to the local crystal structure in the vicinity of vacancies,
they are expected (especially in the limit of infinite supercell size) to give more accurate predictions than the VCA.
However, the supercell method is limited by the much larger computation time required to perform calculations.
This is because the cells have to be large enough that the results are not dominated by vacancy--vacancy
interactions, and because several calculations with vacancies placed in different locations need to be
performed to correctly describe the random positioning of vacancies in a material.
Furthermore, small supercells also restrict the available compositions.

For stoichiometric MgCNi$_{3}$, calculations were performed with a cut-off for plane-waves in the
interstitial region defined by $|{\bf G}+{\bf k}|_{\text{max}}=8.5/R_{\text{MT}}$, where $R_{\text{MT}}$
is the average muffin tin radius.
Convergence was achieved on a $32\times32\times32$ ${\bf k}$-point mesh resulting in 969 ${\bf k}$-points
in the irreducible Brillouin zone.
The VCA calculations used the same parameters as the stoichiometric calculation, except the C and Ni atoms
were replaced with effective atoms with nuclear charges of 5.58 and 26.63, respectively, thereby simulating
the XRD stoichiometry.
For the $2\times2\times2$ supercell calculations, one C and one Ni atom were removed (thereby giving an
effective stoichiometry of MgC$_{0.875}$Ni$_{2.875}$), and a cut-off for plane-waves in the interstitial
region determined by $|{\bf G}+{\bf k}|_{\text{max}}=7.0/R_{\text{MT}}$ was used, with an $8\times8\times8$
${\bf k}$-point mesh resulting in 120 ${\bf k}$-points in the irreducible supercell Brillouin zone,
corresponding to the same effective ${\bf k}$-point density as the stoichiometric and VCA calculations.
For all of the \textsc{Elk} calculations, the muffin tin radii were set to 2.20 a.u., 1.10 a.u., and 2.40 a.u.,
for Mg, C, and Ni, respectively, and the Perdew-Wang/Ceperly-Alder local density approximation (PWCA-LDA)
\cite{perdew:92} was used for the exchange-correlation functional.
For consistency, the ambient temperature perovskite structure (space group $Pm\bar{3}m$), with Wyckoff
positions of Mg at the $1a$ site $(0,0,0)$, C at $1b$ $(0.5,0.5,0.5)$, and Ni at the $3c$ site $(0.5,0.5,0)$,
and the experimental lattice constant for the nearly stoichiometric polycrystals of 3.81~\AA~was used for
all calculations.
This is because the experimental lattice constants at some of the compositions simulated in the calculations
are unknown, and any estimations would likely contain a large uncertainty.
Tests performed for the non-stoichiometric calculations at different lattice constants, including our XRD value,
indicated negligible changes to the electronic structure compared to those introduced by the varying stoichiometry.

For the $2\times2\times2$ supercells used in this study, there are three distinct configurations for the
removal of a single C and Ni atom that are not related by symmetry or translation.
The characteristic quantity of these configurations is the vacancy--vacancy distance in the periodic supercells.
In the supercells, a vacancy separation of 4.25~\AA~appears twice as often by symmetry and translation
as those with separations of 1.90~\AA, and 5.70~\AA.
Test calculations were performed for different configurations with the same vacancy separation and these showed
the same DOS (to within integration errors).
The effect of relaxing the supercell crystal structure with vacancies was investigated for one configuration.
This relaxation was found to have an almost negligible effect on the DOS compared to the effect of the vacancies
themselves, whilst greatly increasing computation time due to the lowered symmetry and resulting increase in the
number of ${\bf k}$-points in the irreducible Brillouin zone.
As such, the unrelaxed crystal structure was used for all supercell calculations.
Therefore, supercells were constructed for each of the three characteristic vacancy--vacancy distances,
and all three were found to present qualitatively similar results.
The weighted configurational average of all of the supercell calculations was taken, and this average
was used for comparisons with the stoichiometric calculations, the non-stoichiometric calculations by
other methods, and the experimental Compton scattering data.

It is worth noting that the Fermi surface of the supercell calculations cannot straightforwardly be obtained in a
way comparable to the stoichiometric and VCA calculations because the crystal lattice has been artificially extended by
including the extra cells.
As a result, the band structure is folded into the supercell Brillouin zone which is much smaller than that of the
stoichiometric and VCA calculations.
However, it is possible to calculate the electron momentum distribution of the supercell calculations, and
fold this distribution into the original Brillouin zone in order to obtain the ${\bf k}$-space occupation
density.
The supercell Fermi surface can then be determined from the configurationally averaged occupation density in the
original Brillouin zone in the same manner that an experimental Fermi surface is obtained from a Compton scattering
experiment.
In order to make comparisons with our Compton scattering measurements, and to determine the Fermi surface from our
supercell calculations, electron momentum densities and Compton profiles were calculated from the calculated electronic
structure by the method of Ernsting {\it et al.} \cite{ernsting:14}.

Calculations were also performed using the Korringa-Kohn-Rostoker (KKR) method with the coherent potential
approximation (CPA) that can effectively treat disorder within a mean field theory \cite{huhne:98}.

\subsection{Compton scattering}

The Compton scattering experiments were performed on beamline BL08W of the SPring-8
synchrotron, Japan. The high-resolution x-ray Compton spectrometer \cite{bl08w}  
was used with an incident x-ray energy of 115~keV and a scattering angle of 165$^{\circ}$.
The spectrometer resolution could be described by a Gaussian with a full-width-at-half-maximum of 0.12~a.u.. 
The measurements were made at room temperature ($T=298$~K), and each profile 
had approximately $10^{5}$ counts in the Compton peak. Corrections were made for 
absorption, analyser and detector efficiencies,
scattering cross-section, double scattering contributions and background.
The core electron contributions, for the composition MgC$_{0.93}$Ni$_{2.85}$ determined by XRD,
were then subtracted from each profile.

Compton scattering is a uniquely powerful probe of the 
ground-state electronic wave function \cite{cooper:85,bansil:01}. Since only
occupied momentum states contribute to the momentum distribution $\rho({\bf{p}})$,  it
can be used for Fermi surface studies.
A Compton profile, $J(p_{z})$, is a double integral of the electron momentum distribution, $\rho({\bf{p}})$,
\begin{equation}
J(p_{z})=\iint\rho({\bf p})~\text{d}p_{x}\text{d}p_{y},
\label{cp}
\end{equation}
where $p_{z}$ is taken along the scattering vector and $\rho({\bf p})$ can be expressed as, 
\begin{equation}
\rho({\bf p})=\sum_{{\bf k},j}n_{{\bf k},j}\bigg|\int\psi_{{\bf k},j}({\bf r})\text{e}^{-\text{i}{\bf p}\cdot{\bf r}}~\text{d}{\bf r}\bigg|^{2},
\end{equation}
where $\psi_{{\bf k},j}({\bf r})$ is the wave function of the electron in band $j$ with wave-vector ${\bf k}$,
and $n_{{\bf k},j}$ is its occupation.

A set of six Compton profiles were measured with scattering vectors equally spaced between the $\Gamma$−-X and $\Gamma$−-M
directions of the simple cubic Brillouin zone. Tomographic reconstruction was used to obtain a once-integrated
momentum distribution (a two-dimensional projection) in the plane of the scattering vectors. Cormack's method
was employed to perform the reconstruction \cite{kontrym-sznajd:90}.
The Lock-Crisp-West prescription \cite{lcw} was then applied to fold the 
projected ${\bf p}$-space distribution back into the first Brillouin zone in order to give the 
projected ${\bf k}$-space occupation density. The Fermi surface, which
separates occupied from unoccupied states in ${\bf k}$-space, is manifest as a
sharp change in the ${\bf k}$-space occupation density.

\section{Acknowledgments}

\begin{acknowledgments}
The Compton scattering experiment was performed with the approval of the
Japan Synchrotron Radiation Research Institute (JASRI, proposal no. 2013B1355). 
Calculations were performed using the computational
facilities of the Advanced Computing Research Centre, University
of Bristol (\verb+http://www.bris.ac.uk/acrc/+).
We acknowledge the financial support of the UK EPSRC (EP/J002925/1).
\end{acknowledgments}

\section{Author Contributions}
S.B.D. initiated the project, with ideas contributed by J.A.D., J.W.T and S.R.G..
N.D.Z. grew the sample and determined the superconducting transition temperature.
H.A.S. performed the x-ray diffraction and refined the structure to determine the stoichiometry.
D.E., D.B., T.E.M., R.A.E., and S.B.D. performed the Compton scattering experiment and D.B., D.E. and S.B.D. analysed the data.
D.E., T.E.M. and S.B.D. performed the electronic structure calculations.
D.B., D.E., S.B.D. and T.E.M. contributed to writing the paper.
This manuscript reflects the contributions and ideas of all authors. 

\section{Additional Information}

\subsection{Competing financial interests.}

The authors declare no competing financial interests.

\subsection{Availability of Materials and Data}

The underlying research materials can be accessed at the following DOI: 10.5523/bris.ulryo0ap77x11zzatcwcgcu5q .

\end{document}